# Ferromagnetism in Correlated Electron Systems: Generalization of Nagaoka's Theorem

Marcus Kollar,* Rainer Strack and Dieter Vollhardt†

*Institut für Theoretische Physik C, RWTH Aachen, 52056 Aachen, Germany*

(November 3, 1995)

## Abstract

Nagaoka's theorem on ferromagnetism in the Hubbard model with one electron less than half filling is generalized to the case where all possible nearest-neighbor Coulomb interactions (the density-density interaction $V$, bond-charge interaction $X$, exchange interaction $F$, and hopping of double occupancies $F'$) are included. It is shown that for ferromagnetic exchange coupling ($F > 0$) ground states with maximum spin are stable already at finite Hubbard interaction $U > U_c$. For non-bipartite lattices this requires a hopping amplitude $t \leq 0$. For vanishing $F$ one obtains $U_c \to \infty$ as in Nagaoka's theorem. This shows that the exchange interaction $F$ is important for stabilizing ferromagnetism at finite $U$. Only in the special case $X = t$ the ferromagnetic state is stable even for $F = 0$, provided the lattice allows the hole to move around loops.

71.27+a, 75.10.Lp

Typeset using REVTEX

# I. INTRODUCTION

The single-band Hubbard model was originally introduced as a model for ferromagnetism of itinerant electrons.[1-3] The model is given by

$$\hat{H}_{Hubbard} = -t \sum_{\langle ij \rangle \sigma} (\hat{c}_{i\sigma}^+ \hat{c}_{j\sigma} + \text{h.c.}) + U \sum_i \hat{n}_{i\uparrow} \hat{n}_{i\downarrow} \qquad (1)$$

where $\hat{c}_{i\sigma}^+$ ($c_{i\sigma}$) creates (annihilates) an electron with spin $\sigma = \uparrow, \downarrow$, $\hat{n}_{i\sigma} = \hat{c}_{i\sigma}^+ \hat{c}_{i\sigma}$ is the number operator, and $\langle ij \rangle$ denotes nearest neighbors. This is the simplest possible correlation model for electrons on a lattice. However, rigorous evidence for itinerant ferromagnetism in this model is very limited. One of the most important results is Nagaoka's theorem,[4] which states that if the Hubbard repulsion $U$ is infinite, the ground state has maximum total spin on certain lattices in the case of precisely one hole. The physical mechanism behind Nagaoka's theorem is the following. If $U = \infty$, the ground state of (1) is macroscopically degenerate. This degeneracy is lifted by the motion of the hole since it is energetically favorable for it to move in a background of fully aligned spins (provided the lattice allows for motion of the hole around loops[4]). A simpler proof of Nagaoka's theorem was later given by Tasaki,[5] who also showed that additional density-dependent interactions do not alter this result.

Several other mechanisms leading to ferromagnetism in the Hubbard model have been discussed since then.[6] Lieb[7] proved that the ground state is ferromagnetic for bipartite lattices with different numbers of sites in each sublattice. Mielke[8] and Tasaki[9] proved the stability of ferromagnetism for special lattices with flat bands.[10] Recently, a new route was taken by Müller-Hartmann[11] who studied ferromagnetism at low particle density in dimension $d = 1$. He included next-nearest neighbor hopping in such a way that the band has two minima. At low density, the on-site repulsion $U$ generates a ferromagnetic exchange coupling between particles in these two pockets.

Clearly, it is still a long way to a true understanding of itinerant ferromagnetism in solids. It is quite obvious that the single-band Hubbard model is not a *generic* model for ferromagnetism. So far, either the assumption of a special kind of hopping, or of $U = \infty$, or both, were necessary to prove the stability of ferromagnetism. One may therefore ask if there exist other, simple mechanisms leading to itinerant ferromagnetism that are not contained in the Hubbard model. There are two important candidates: (i) band degeneracy, as it exists in $3d$-transition metals, and (ii) nearest-neighbor exchange interaction, which is always present in a fermionic system with Coulomb interaction. Here we discuss only the latter, since the effect of band degeneracy will be discussed separately.[12]

From atomic magnetism it is known that there exists a ferromagnetic Hund's rule interaction between orbitals on the same atom. A ferromagnetic spin wave function is symmetric under exchange of particles, and for the electron wave function to be overall antisymmetric its coordinate part must be antisymmetric itself. An antisymmetric coordinate wave function is zero whenever two particles are at the same position and thus minimizes the repulsive Coulomb interaction. This is the well-known mechanism leading to the first Hund's rule for atomic magnetism. In solids a similar ferromagnetic Heisenberg exchange term is present even between orbitals at different sites. However, since their overlap is much smaller than for orbitals on the same atom, this interaction may be quite small. Nonetheless, it will not be strictly zero. Therefore this direct exchange interaction, denoted by $F$ below, provides a natural way for stabilizing ferromagnetic states.[13] Of course, other features of the model, in particular the hopping $t$ and the structure of the lattice, are also important factors concerning the stability of ferromagnetism in the ground state.

With this in mind it is worthwhile to review the steps that originally led to the Hubbard model, and to retain, in a systematic way, Coulomb interaction terms beyond the on-site repulsion $U$. This is done in Sec. II, and a model Hamiltonian with all nearest neighbor interactions is derived. In Sec. III we state sufficient stability conditions for ferromagnetic ground states in the case of one hole in a half-filled band. In particular, it turns out that if the direct exchange is ferromagnetic ($F > 0$), and even if $F = 0$ in a special case, the on-site repulsion $U$ need only be larger than a finite value $U_c$, thereby generalizing Nagaoka's theorem to finite $U$. The details of the proof, using a method employed previously for the case of half-filling,[14] are deferred to the Appendix. Sec. IV contains our conclusions.

## II. DERIVATION OF THE MODEL

Let us first review the derivation of effective models for metallic ferromagnetism. The general electronic model expected to describe ferromagnetic phase transitions in transition metals was introduced by Hubbard.[1] It is given by the electronic Hamiltonian[15]

$$\hat{H} = \sum_{ij} t_{ij}^{\alpha} \hat{c}_{i\alpha\sigma}^{+} \hat{c}_{j\alpha\sigma} + \sum_{ijmn} v_{ijmn}^{\alpha\beta\mu\nu} \hat{c}_{i\alpha\sigma}^{+} \hat{c}_{j\beta\sigma'}^{+} \hat{c}_{n\nu\sigma'} \hat{c}_{m\mu\sigma} \qquad (2)$$

Here, $\hat{c}_{i\alpha\sigma}^{+}$ ($\hat{c}_{i\alpha\sigma}$) creates (annihilates) an electron with spin $\sigma$ in a Wannier orbital $\alpha$ localized at site $i$. The first term describes hopping between two sites $i$, $j$ and contains the kinetic energy and the ionic potential $U_{ion}(\mathbf{r})$. The second term describes the (screened) Coulomb interaction between electrons, $V_{ee}(\mathbf{r} - \mathbf{r}')$.[16] The matrix elements, expressed in the Wannier basis, are ($\hbar \equiv 1$)

$$t_{ij}^{\alpha} = \langle i\alpha | -\frac{1}{2m}\nabla^2 + U_{ion}(\mathbf{r}) | j\alpha \rangle \qquad (3a)$$

$$v_{ijmn}^{\alpha\beta\mu\nu} = \langle i\alpha, j\beta | V_{ee}(\mathbf{r} - \mathbf{r}') | m\mu, n\nu \rangle \qquad (3b)$$

So far no approximation was made. The Hamiltonian (2) contains infinitely many parameters. For simplicity it is therefore often assumed that the essential physics of the problem is captured by a single $s$-band, whereby all other bands are neglected. More precisely, all other bands are *projected* onto one single *effective* $s$-band. This approximation requires the existence of a band gap above the effective band. Then the deviation of the parameters $t_{ij}$ and $v_{ijmn}$ from their multi-band values can be determined, in principle, by perturbation theory.

The restriction to a single $s$-band entails considerable simplifications: orbital indices may be dropped in eq. (2) and (3); furthermore, all matrix elements depend only on the separation of the lattice sites (and not on direction). Since the matrix elements are expected to fall off quickly with distance, one usually retains only the first few of them. Thus hopping is restricted to nearest neighbor sites $i$ and $j$: $-t \equiv t_{ij}$. It is also natural to assume that $U \equiv v_{iiii}$ is the largest matrix element of the Coulomb interaction. Keeping only $t$ and $U$ one obtains the Hubbard model, eq. (1).

However, there are other terms that can be of appreciable size.[1] These are the *two-site* terms of the interaction: $V \equiv v_{ijij}$, $X \equiv v_{iiij}$, $F \equiv v_{ijji}$, $F' \equiv v_{iijj}$, where $i$ and $j$ are nearest neighbors. Keeping these terms one obtains the following single-band model:

$$\hat{H}_{NN} = \hat{H}_{Hubbard} + V \sum_{\langle ij \rangle} \hat{n}_i \hat{n}_j + X \sum_{\langle ij \rangle \sigma} (\hat{c}^+_{i\sigma} \hat{c}_{j\sigma} + \text{h.c.})(\hat{n}_{i-\sigma} + \hat{n}_{j-\sigma})$$
$$+ F \sum_{\langle ij \rangle \sigma \sigma'} \hat{c}^+_{i\sigma} \hat{c}^+_{j\sigma'} \hat{c}_{i\sigma'} \hat{c}_{j\sigma} + F' \sum_{\langle ij \rangle} (\hat{c}^+_{i\uparrow} \hat{c}^+_{i\downarrow} \hat{c}_{j\downarrow} \hat{c}_{j\uparrow} + \text{h.c.}). \quad (4)$$

Here $V$ is the density-density interaction between nearest neighbors, $X$ is the bond-charge interaction giving rise to correlated hopping, $F$ is the exchange interaction discussed in the introduction (ferromagnetic in nature if $F > 0$), and $F'$ represents hopping of double occupancies.

While the on-site interaction $U$ usually has the largest numerical value, the other matrix elements are certainly not zero. Hubbard's estimates[1] for transition metals are, for example, $U \approx 10 eV$, $V \approx 2\text{-}3 eV$, $X \approx 1 eV$ and $F, F' \approx \frac{1}{40} eV$, and the hopping amplitude $t$ typically ranges between $0.5 eV$ and $1.5 eV$. Even if nearest neighbor interactions *are* very small, they can be qualitatively important if they have different symmetries than the $U$-term and thus can lift degeneracies.

The model (4) was essentially derived already by Hubbard.[1] Extensions of the actual Hubbard model (1) by some or all of the terms in eq. (4) received much attention since then. For example, Campbell, Gammel and Low[17] presented a detailed investigation of the phase diagram of $\hat{H}_{NN}$ in dimension $d = 1$, and discussed the relative magnitude of its parameters for real materials. On a mean-field level, the effect of the terms in (4) on the stability of ferromagnetism was studied by Hirsch.[18] Furthermore, exact solutions are possible in the special case of $X = t$. In this case the number of doubly occupied sites is a conserved quantity, and the exact ground state solution can be obtained in a wide range of parameters.[19–22] For $X = t$ and $V = F = F' = 0$ the model was recently solved exactly in one dimension,[23,24] while for $X = t = -V = F = F'$ a solvable supersymmetric model is obtained.[25,26] The case $X = t$ will play a special role in our analysis, too.

Criteria for the stability of ferromagnetic ground states of the Hamiltonian (4) were recently derived for the case of half-filling (one electron per site).[14,19–22] The ferromagnetic states are then found to be insulating. To gain insight into the more general problem of *itinerant* ferromagnetism we will now investigate a half-filled band *with one hole*, as in Nagaoka's work.[4] Thus we consider a finite lattice with $L$ sites and fix the total number of particles at $N = L - 1$. The number of nearest neighbors is denoted by $Z$. We consider lattices with at least $Z$ nearest neighbor bonds between any subset of lattice sites and the set of remaining sites. For example, all crystal lattices with periodic boundary conditions fulfill this requirement.

### III. FERROMAGNETIC GROUND STATES

The Hamiltonian $\hat{H}_{NN}$ commutes with the total spin $\hat{\mathbf{S}} = \sum_i \hat{\mathbf{S}}_i$, where $\hat{\mathbf{S}}_i = \frac{1}{2} \sum_{\sigma \sigma'} \hat{c}^+_{i\sigma} \boldsymbol{\tau}_{\sigma \sigma'} \hat{c}_{i\sigma'}$ and $\boldsymbol{\tau}$ are the Pauli matrices. The eigenvalues of $\hat{\mathbf{S}}^2$ are denoted by $S(S+1)$. In the following we will be concerned only with saturated ferromagnetic states with largest possible eigenvalue $S_{max} \equiv N/2 = (L-1)/2$. There are $2S_{max} + 1 = L$ such states with the same energy eigenvalue.

We are interested in the following question: *Under which circumstances do the ground states of $\hat{H}_{NN}$ have maximum spin?* For the pure Hubbard model (i. e, $V = X = F = F' = 0$), Nagaoka's theorem[4] states that for $U = \infty$, $t < 0$ ($t \neq 0$ if the lattice is bipartite) the ground states have $S = S_{max}$. This statement can be generalized to arbitrary density-density interaction $V$.[5] These results require the lattice to have "loops" as discussed in the introduction. For example, this is the case for the square, triangular, simple cubic, body-centered cubic, face-centered cubic, and hexagonal close packed lattice, but not in one dimension or on the Bethe lattice.[4]

The main result of this paper is a generalization of Nagaoka's theorem to finite values of the Hubbard interaction $U$. In the Appendix we derive the following conditions for ferromagnetic ground states:

*The ground states of $\hat{H}_{NN}$ with one hole (i. e. $N = L-1$) have maximum total spin $S = S_{max} = \frac{L-1}{2}$ in the following cases:*

*Case 1: On any lattice, if $F > 0$, $t \leq 0$ and (a) $X \neq t$ and $U > U_c^{(1)}$, or (b) $X = t$ and $U \geq U_c^{(2)}$.*

*Case 2: On lattices with loops, if $X = t < 0$, $F = 0$, and $U > U_c^{(2)}$.*

*In both cases $t > 0$ is allowed if the lattice is bipartite.*

These results are summarized in Table I. The constants $U_c^{(1)}$ and $U_c^{(2)}$ are given by

$$U_c^{(1)} = Z\left(2|t| + \left|V - F - 2|t|\right| + \frac{(X-t)^2}{F} + \left|F' - \frac{(X-t)^2}{F}\right|\right), \tag{5a}$$

$$U_c^{(2)} = Z\left(2|t| + \left|V - \frac{F}{2} - 2|t|\right| + |F'|\right). \tag{5b}$$

Hence, if $F > 0$, ferromagnetic ground states are stable on any lattice for $U$ larger than a *finite* critical value. For $F \to 0^+$ we have $U_c^{(1)} \to \infty$, thus yielding Nagaoka's condition for the pure Hubbard model. This shows that the Heisenberg interaction $F$, which is neglected in the Hubbard model, provides an obvious mechanism for stabilizing ferromagnetic ground states at finite $U$. Note that since $X$ and $t$ are expected to be of the same order of magnitude, the sensitive dependence on $F$, due to the term $\frac{(X-t)^2}{F}$, may cancel from $U_c^{(1)}$, and values of the order of $U_c \sim 12 eV$ are possible. The dependence of $U_c$ on $t, V, F$ is depicted in Fig. 1. The case $X = t$ is special, since in this case the stability of ferromagnetism can be achieved either by $F > 0$, or by $F \geq 0$ and $t < 0$ if the lattice has loops.

| Case | | Condition on $U$ | Condition on lattice | Condition on $t$ |
|---|---|---|---|---|
| 1a | $F > 0$, $X \neq t$ | $U > U_c^{(1)}$ | any lattice | bipartite lattice: $t$ arbitrary<br>non-bipartite lattice: $t \leq 0$ |
| 1b | $F > 0$, $X = t$ | $U \geq U_c^{(2)}$ | any lattice | bipartite lattice: $t$ arbitrary<br>non-bipartite lattice: $t \leq 0$ |
| 2 | $F = 0$, $X = t$ | $U > U_c^{(2)}$ | lattice with loops | bipartite lattice: $t \neq 0$<br>non-bipartite lattice: $t < 0$ |

TABLE I. Sufficient conditions for ferromagnetic ground states with one hole.

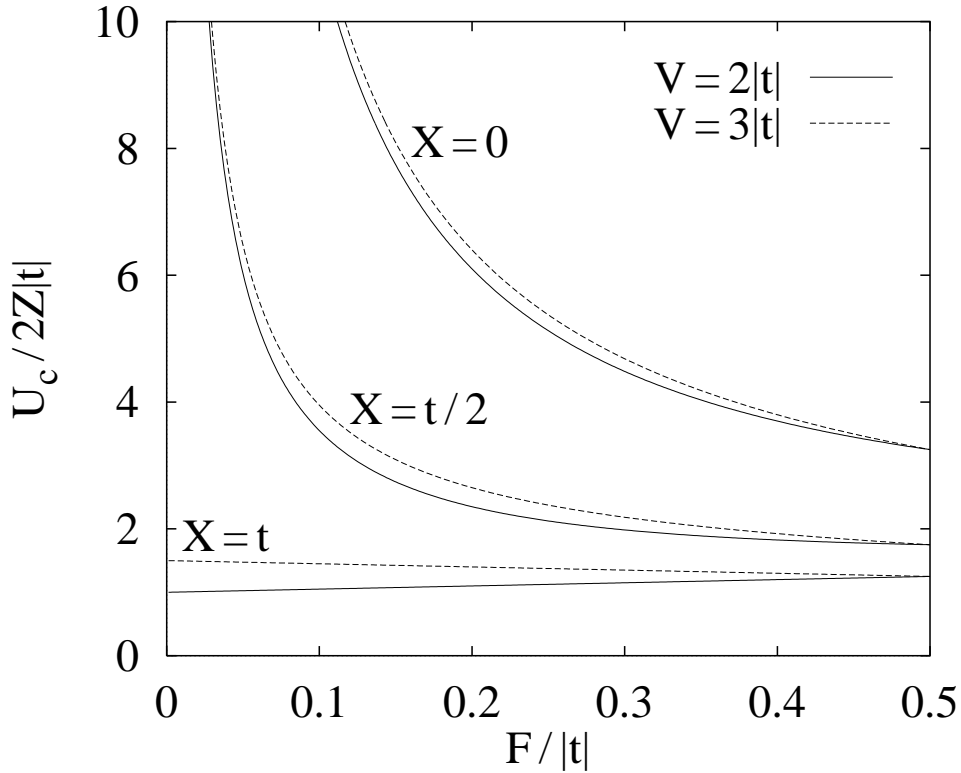

FIG. 1. Critical value $U_c$ vs. exchange interaction $F$ for different $t$, $V$, $X$, and $F' = 0$. For $U > U_c$ the ground state is ferromagnetic. See Table I for details.

The critical couplings $U_c^{(1)}$ and $U_c^{(2)}$ are sums of terms, each of which corresponds to a typical energy scale. This means that the on-site interaction $U$ has to be larger than the energy describing the paramagnetic state (bandwidth $\sim Z|t|$), as well as the threshold energies for the onset of a charge-density wave or phase separation ($\sim Z|V|$), $\eta$-pairing superconductivity[21] ($\sim Z|F'|$), and a spin-density wave ($\sim (X-t)^2/F$). Note, however, that these terms do not enter separately, but appear in combinations, i. e. the effects interfere as should be expected.

We remark that the above conditions are *sufficient* conditions. The occurrence of ground states with maximum spin outside the above parameter region is not ruled out. For example, it may still be possible to find ferromagnetic ground states even for $X \neq t$, $F = 0$ and $U_c < \infty$. It is interesting to note that the values for $U_c$ appearing in eq. (5) are the same as those for the case of half-filling (no hole).[14] The bound $U_c$ for half-filling was recently improved by Schadschneider and de Boer.[22]

As shown in the Appendix, the ferromagnetic ground states are the same as those discussed by Nagaoka,[4] i. e. the wave function with $\hat{S}^z = S_{max}$ corresponds to a band filled with spin-up electrons, with the hole at the top of the band, as illustrated in Fig. 2. If $t < 0$ the band maximum is at the origin, and the corresponding wave function is $|\psi_0\rangle = \hat{a}_{0\uparrow}|\uparrow\rangle$, where $\hat{a}_{\mathbf{k}\sigma} = \frac{1}{\sqrt{L}} \sum_i \exp(i\mathbf{k}\mathbf{R}_i)\hat{c}_{i\sigma}$, and $|\uparrow\rangle = \prod_i \hat{c}_{i\uparrow}^+|0\rangle = \prod_{\mathbf{k}} \hat{a}_{\mathbf{k}\uparrow}^+|0\rangle$ is the filled spin-up band. For $t > 0$ a bipartite lattice is required and the band maximum is at wave vector $\mathbf{Q}$ defined by $\epsilon_{\mathbf{k}+\mathbf{Q}} = -\epsilon_{\mathbf{k}}$, e. g. for a hypercubic lattice we have $\mathbf{Q} = (\pi, \pi, \ldots \pi)$. In this case the wave function is $|\psi_0\rangle = \hat{a}_{\mathbf{Q}\uparrow}|\uparrow\rangle$. All $2S_{max} + 1$ ground states can be obtained from $|\psi_0\rangle$ by global

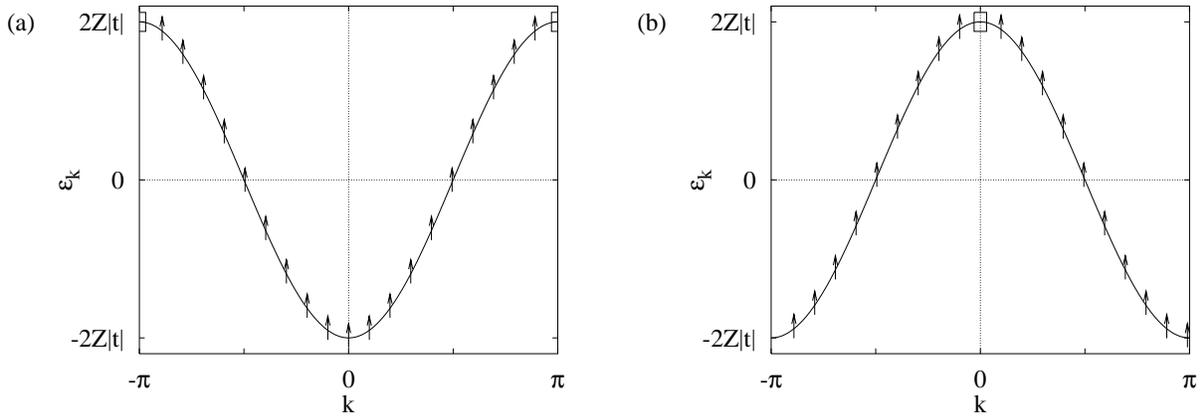

FIG. 2. The figure illustrates (for $d = 1$) that the ferromagnetic ground state corresponds to a band filled with spin-up electrons (indicated by arrows), from which a single electron has been removed at the top of the band (open square). (a) $t > 0$ (with a hole at $k = \pi$); (b) $t < 0$ (with a hole at $k = 0$).

$SU(2)$ rotations, i. e. the state

$$|\psi_M\rangle = (\hat{S}^x - i\hat{S}^y)^M |\psi_0\rangle \qquad (6)$$

has $\hat{S}^z = S_{max} - M$, where $M = 0 \ldots L - 1$. The states $\{|\psi_M\rangle\}$ have the same energy eigenvalue.

If $F > 0$, our criteria permit ferromagnetism in any spatial dimension. In particular they hold in dimension $d = 1$. This is not in conflict with the Lieb-Mattis theorem[27] which rules out ferromagnetism for any continuum model with $k^2$-dispersion and symmetric interaction potential in $d = 1$. In a Bloch basis a $k^2$-dispersion corresponds to infinitely many bands as in (2), whereas the Hamiltonian $\hat{H}_{NN}$ in (4) is obtained by a projection onto *one* of these bands, as discussed in Sec. II. Hence the model (4) is quite different from the initial multi-band Hamiltonian $\hat{H}$ in (2), in particular there does not exist a potential energy in a continuum model that gives rise to exactly the terms appearing in the truncated Hamiltonian $\hat{H}_{NN}$. Therefore the Lieb-Mattis theorem does not apply to the single-band Hamiltonian (4) in $d = 1$. In the Appendix of Ref. 27 Lieb and Mattis proved a second theorem, using the occupation number formalism, which precludes ferromagnetism in $d = 1$ in the case of next-neighbor hopping and purely density-dependent (i. e. site-diagonal) interactions. Therefore this second theorem also does not apply to the Hamiltonian $\hat{H}_{NN}$, eq. (4), where the *non-diagonal* exchange interaction $F > 0$ (which is always present as part of the Coulomb interaction) plays a crucial role. The role of non-diagonal interactions for the stability of ferromagnetism was already discussed by these authors.[27]

## IV. CONCLUSIONS

We presented a generalization of Nagaoka's theorem to a Hubbard model with all nearest-neighbor interactions. For this model, with one hole in a half-filled band, we derived rigorous, sufficient conditions for the stability of saturated ferromagnetism in the ground state. The ferromagnetic ground state is found to be stable on any lattice provided the next-neighbor exchange $F$ is ferromagnetic and the Hubbard repulsion $U$ is larger than a critical value $U_c < \infty$, with $U_c \sim 1/F$ for $F \to 0^+$. If $F = 0$, only the special case $X = t$ (unlikely to be fulfilled exactly in real materials) yields a ferromagnetic ground state on lattices with loops.

The ferromagnetic ground state is an itinerant state with non-zero kinetic energy. The proof of its stability cannot be easily extended to doping beyond a single hole. Of course, this would be highly desirable since a single hole is irrelevant in the thermodynamic limit. However, the ground states of the model with next neighbor interactions are difficult to obtain for finite hole densities, since in this case simple eigenstates of the Hamiltonian are not known, such that the present methods can be applied only in special cases.[12]

To be able to explain ferromagnetism in more detail, it is clear that more ingredients are needed than those contained in the single-band Hubbard model with nearest neighbor hopping. Of greatest interest is the case of band degeneracy where the present methods can be applied, too.[12]

The authors would like to acknowledge valuable discussions with D. Baeriswyl, E. H. Lieb, D. C. Mattis, E. Müller-Hartmann, and A. Schadschneider. M.K. and R.S. gratefully acknowledge scholarships of the Studienstiftung des Deutschen Volkes. This work was supported in part by the Sonderforschungsbereich 341 of the Deutsche Forschungsgemeinschaft.

## APPENDIX: DETAILS OF THE DERIVATION

To derive the sufficient conditions for ferromagnetic ground states stated in Sec. III we rearrange the Hamiltonian (4) as a sum of positive semi-definite terms. This enables us to construct a lower bound on the ground state energy. If this lower bound coincides with the eigenvalue of a trial state, this state is an exact ground state of (4).[28,29] Here we consider a wave function of the form

$$|\psi_0\rangle = \sum_j a_j \hat{c}_{j\uparrow} |\uparrow\rangle,$$

where $|\uparrow\rangle = \prod_i \hat{c}_{i\uparrow}^+ |0\rangle$ is the state with all sites occupied by spin-up electrons. The coefficients $a_j \neq 0$ will be determined below. $|\psi_0\rangle$ is a state with $N = L - 1$ particles. Since it is an eigenstate of $\hat{S}^z$ with eigenvalue $S_{max}$, it is a representative of the subspace with $S = S_{max}$. It is sufficient to consider a state of this form, because whenever this state is a ground state, so are its $(2S + 1) = L$ global $SU(2)$ rotations.

We introduce the operators

$$\hat{P}_{ij\sigma} = (1 - \hat{n}_{i-\sigma})(\hat{c}_{i\sigma} + \lambda_1 \hat{c}_{j\sigma})(1 - \hat{n}_{j-\sigma})$$
$$\hat{Q}_{ij\sigma} = \hat{n}_{i-\sigma}(\hat{c}_{i\sigma} + \lambda_1 \hat{c}_{j\sigma})\hat{n}_{j-\sigma},$$
$$\hat{A}_{ij} = \alpha^{-1}(\hat{c}_{i\downarrow}\hat{c}_{i\uparrow} + \hat{c}_{j\downarrow}\hat{c}_{j\uparrow}) + \alpha\lambda_2(\hat{c}_{j\downarrow}\hat{c}_{i\uparrow} + \hat{c}_{i\downarrow}\hat{c}_{j\uparrow}),$$
$$\hat{B}_{ij} = \hat{c}_{i\downarrow}\hat{c}_{i\uparrow} + \lambda_3 \hat{c}_{j\downarrow}\hat{c}_{j\uparrow}.$$

Here $\lambda_1 = \mathrm{sgn}(t)$, $\lambda_2 = \mathrm{sgn}(X - t)$, $\lambda_3 = \mathrm{sgn}(F' - |X - t|/\alpha^2)$, and $\alpha \neq 0$ is an arbitrary, real parameter. Furthermore we introduce

$$\hat{\Omega}_{ij} = \begin{cases} \hat{e}_i \hat{e}_j + 3\hat{d}_i \hat{d}_j + \hat{p}_i \hat{d}_j + \hat{p}_j \hat{d}_i & \text{if } V > 2|t| + (F + \alpha^2 |X - t|)/2 \\ \frac{1}{2}(\hat{p}_i - \hat{p}_j)^2 + 2(\hat{d}_i \hat{e}_j + \hat{d}_j \hat{e}_i) & \text{if } V < 2|t| + (F + \alpha^2 |X - t|)/2 \end{cases},$$

where $\hat{e}_i = (1 - \hat{n}_{i\uparrow})(1 - \hat{n}_{i\downarrow})$, $\hat{p}_i = (\hat{n}_{i\uparrow} - \hat{n}_{i\downarrow})^2$, $\hat{d}_i = \hat{n}_{i\uparrow}\hat{n}_{i\downarrow}$ are the projectors onto an empty, singly and doubly occupied site, respectively. It is straightforward to verify that $\hat{H}_{NN}$ can be written as

$$\hat{H}_{NN} = \sum_{\langle ij \rangle} \left[ |t| \sum_{\sigma} (\hat{P}_{ij\sigma} \hat{P}^+_{ij\sigma} + \hat{Q}^+_{ij\sigma} \hat{Q}_{ij\sigma}) + |X - t| \hat{A}^+_{ij} \hat{A}_{ij} + |\widetilde{F}'| \hat{B}^+_{ij} \hat{B}_{ij} \right]$$
$$+ |\widetilde{V}| \sum_{\langle ij \rangle} \hat{\Omega}_{ij} + \widetilde{U} \sum_i \hat{n}_{i\uparrow} \hat{n}_{i\downarrow} - 2\widetilde{F} \sum_{\langle ij \rangle} \hat{\mathbf{S}}_i \hat{\mathbf{S}}_j, \tag{A1}$$

where

$$\widetilde{F}' = F' - \frac{|X - t|}{\alpha^2}, \tag{A2a}$$

$$\widetilde{F} = F - \alpha^2 |X - t|, \tag{A2b}$$

$$\widetilde{V} = V - \frac{F + \alpha^2 |X - t|}{2} - 2|t|, \tag{A2c}$$

$$\widetilde{U} = U - Z\left(2|t| + |\widetilde{V}| + \frac{|X - t|}{\alpha^2} + |\widetilde{F}'|\right) \tag{A2d}$$

and an overall constant was dropped.

Let us consider the terms in (A1) one by one. The first term is positive semi-definite. Since $\hat{A}_{ij}|\psi_0\rangle = 0$, $\hat{B}_{ij}|\psi_0\rangle = 0$, and $\hat{Q}_{ij\sigma}|\psi_0\rangle = 0$, the action of this term on $|\psi_0\rangle$ is given by

$$|t| \sum_{\langle ij \rangle \sigma} \hat{P}_{ij\sigma} \hat{P}^+_{ij\sigma} |\psi_0\rangle = |t| \sum_{\langle ij \rangle m\sigma} (1 - \hat{n}_{i-\sigma})(1 - \hat{n}_{j-\sigma})(\hat{c}_{i\sigma} + \lambda_1 \hat{c}_{j\sigma})(\hat{c}^+_{i\sigma} + \lambda_1 \hat{c}^+_{j\sigma}) a_l \hat{c}_{m\uparrow} |\Uparrow\rangle$$
$$= |t| \sum_{\langle ij \rangle \sigma} (\hat{c}_{i\uparrow} + \lambda_1 \hat{c}_{j\uparrow})(\hat{c}^+_{i\uparrow} + \lambda_1 \hat{c}^+_{j\uparrow})(a_i \hat{c}^+_{i\uparrow} + a_j \hat{c}^+_{j\uparrow}) |\Uparrow\rangle$$
$$= |t| \sum_{\langle ij \rangle \sigma} (a_i + \lambda_1 a_j)(\hat{c}_{i\uparrow} + \lambda_1 \hat{c}_{j\uparrow}) |\Uparrow\rangle.$$

Thus, if $a_i = -\text{sgn}(t) a_j$ for all nearest neighbors $\langle ij \rangle$, we find that $|\psi_0\rangle$ has zero eigenvalue and hence is a ground state of this term. Let us consider the case $t < 0$ first. Then we must have $a_i = a_j$, and the resulting wave function is

$$|\psi_0\rangle = \frac{1}{\sqrt{L}} \sum_i \hat{c}_{i\uparrow} |\Uparrow\rangle = \hat{a}_{\mathbf{k}=0\uparrow} |\Uparrow\rangle,$$

where $\hat{a}_{\mathbf{k}\sigma} = \frac{1}{\sqrt{L}} \sum_i \exp(i\mathbf{k}\mathbf{R}_i) \hat{c}_{i\sigma}$. Note that the filled spin-up band can be expressed as $|\Uparrow\rangle = \prod_{\mathbf{k}} \hat{a}_{\mathbf{k}\uparrow} |0\rangle$. The (non-interacting) band structure is given by $\epsilon_{\mathbf{k}} = -t \sum_{\langle ij \rangle} \exp(i\mathbf{k}(\mathbf{R}_i - \mathbf{R}_j))$. Thus for $t < 0$ the hole at $k = 0$ is created at the *band maximum*, where $\epsilon_{\mathbf{k}=0} = Z|t|$. This is shown in Fig. 2a.

Next consider $t > 0$. Then $a_i = -a_j$ must hold for all nearest neighbors $\langle ij \rangle$. This is only possible if the lattice is bipartite, i. e. if any two nearest neighbors are located on different sublattices $\mathcal{A}$ and $\mathcal{B}$. In this case the wave function becomes

$$|\psi_0\rangle = \frac{1}{\sqrt{L}}(\sum_{i\in\mathcal{A}} \hat{c}_{i\uparrow} - \sum_{i\in\mathcal{B}} \hat{c}_{i\uparrow})|\uparrow\rangle = \hat{a}_{\mathbf{k}=\mathbf{Q}\uparrow}|\uparrow\rangle. \tag{A3}$$

Here $\mathbf{Q}$ is defined by $\exp(i\mathbf{Q}(\mathbf{R}_i - \mathbf{R}_j)) = -1$ for nearest neighbors $\langle ij\rangle$, or equivalently by $\epsilon_{\mathbf{k}+\mathbf{Q}} = -\epsilon_{\mathbf{k}}$. This implies that the band maximum is at $\mathbf{k} = \mathbf{Q}$ for $t > 0$, and this is the state in which the hole is created, as shown in Fig. 2b.

From now on we choose the coefficients $a_i$ as just described, $|\psi_0\rangle$ thus being a ground state of the first term in (A1). If the lattice is not bipartite, this requires $t \leq 0$. (The trivial case $t = 0$ can also be included here.)

Turning to the positive semi-definite $\Omega$-term in (A1), we observe that for $\widetilde{V} \geq 0$ it annihilates $|\psi_0\rangle$, which in this case is a ground state of this term. Now consider $\widetilde{V} < 0$, in which case $|\psi_0\rangle$ is an eigenstate of $\sum_{\langle ij\rangle}\hat{\Omega}_{ij}$ with eigenvalue $Z/2$. We now show that this is the lowest possible eigenvalue for $N = L - 1$. Consider states with a fixed configuration $\{p_i = 0, 1\}$ of $P$ singly occupied sites, $P = \sum_i p_i$. The possible (integer) values of $P$ are $0 \leq P \leq L$.

(i) $P = L$. This case is impossible for $N = L - 1$.

(ii) $1 \leq P \leq L - 1$. In this case we have

$$\Omega \equiv \langle\sum_{\langle ij\rangle}\hat{\Omega}_{ij}\rangle \geq \frac{1}{2}\sum_{\langle ij\rangle}(p_i - p_j)^2 = \frac{1}{2}\sum_{\langle ij\rangle}(1 - \delta_{p_i p_j}),$$

i. e. $2\Omega$ is bound from below by the number of bonds $\langle ij\rangle$ with $p_i \neq p_j$. Consider the set of sites with $p_i = 1$. By our definition of a lattice at the end of Sec. II, these sites are connected to at least $Z$ sites with $p_j = 0$. Hence $\Omega \geq Z/2$.

(iii) $P = 0$. Then, since $N = L - 1$, the number of lattice sites $L$ must necessarily be odd, and there must be $\frac{L+1}{2}$ empty and $\frac{L-1}{2}$ doubly occupied sites. (There are no singly occupied sites.) Consider states with a fixed configuration of doubly occupied sites $\{d_i = 0, 1\}$. We have

$$\Omega \geq 2\sum_{\langle ij\rangle}(d_i(1-d_j) + d_j(1-d_i)) = 2\sum_{\langle ij\rangle}(1 - \delta_{d_i d_j}).$$

Except for a factor of 4, this is the same problem as above, with $\{p_i\}$ replaced by $\{d_i\}$. Therefore in this case $\Omega \geq 2Z > Z/2$.

Thus, summarizing the cases (i)-(iii), we obtain $\Omega \geq Z/2$. Hence $|\psi_0\rangle$ is a ground state of the $\Omega$-term for any $\widetilde{V}$, since it is always an eigenstate with the lowest possible eigenvalue.

Finally, $|\psi_0\rangle$ is clearly a ground state of the remaining terms in (A1) if $\widetilde{U} \geq 0$ and $\widetilde{F} \geq 0$, since it has no doubly occupied sites and maximum spin.

So far we proved that $|\psi_0\rangle$ and its global $SU(2)$ rotations are among the ground states of $\hat{H}_{NN}$ if $\widetilde{U} \geq 0$ and $\widetilde{F} \geq 0$ (and $t \leq 0$ if the lattice is not bipartite). Clearly, if $\widetilde{F} > 0$ (and $\widetilde{U} \geq 0$) these are the only ground states, since then only states with maximum spin minimize the $\widetilde{F}$-term. This will be used below, when we prove that *all* ground states of $\hat{H}_{NN}$ have $S = S_{max}$ in the cases listed in Table I.

*Case 1a:* $X \neq t$, $F > 0$, $U > U_c^{(1)}$. In this case we choose $\alpha^2 \equiv (1-\epsilon)F/|X - t| > 0$ with $0 < \epsilon \leq \frac{1}{2}$ to be specified later. Then $\widetilde{U}$ in eq. (A2) becomes

$$\tilde{U} = U - Z\left(2|t| + \left|V - F - 2|t|\right| + \frac{\epsilon F}{2}\right| + \max(F', \frac{2(X-t)^2}{(1-\epsilon)F} - F')\right).$$

Using $(1-\epsilon)^{-1} \geq 1 + 2\epsilon$ we obtain the bound

$$\tilde{U} \geq U - Z\left(2|t| + \left|V - F - 2|t|\right| + \max(F', \frac{2(X-t)^2}{F} - F') + \epsilon(\frac{F}{2} + \frac{4(X-t)^2}{F})\right)$$
$$= U - U_c^{(1)} - \frac{\epsilon Z}{2F}(F^2 + 8(X-t)^2), \tag{A4}$$

where $U_c^{(1)}$ is defined in eq. (5). Since $U - U_c^{(1)} > 0$ by assumption, it can be seen from eq. (A4) that $\tilde{U}$ is positive if $\epsilon$ is chosen small enough. For example,

$$\epsilon \equiv \min\left(\frac{F(U - U_c^{(1)})}{Z(F^2 + 8(X-t)^2)}, \frac{1}{2}\right)$$

indeed yields $\tilde{U} > 0$. Furthermore, from eq. (A2), $\tilde{F} = \epsilon F > 0$. Hence the present choice of $\alpha^2$ yields $\tilde{U} > 0$ and $\tilde{F} > 0$. Therefore only states with $S = S_{max}$ are ground states of $\hat{H}_{NN}$.

*Case 1b:* $X = t$, $F > 0$, $U \geq U_c^{(2)}$, with $U_c^{(2)}$ as defined in eq. (5). In this case $\tilde{U} = U - U_c^{(2)} \geq 0$ and $\tilde{F} = F > 0$. Again, only states with $S = S_{max}$ are ground states of $\hat{H}_{NN}$.

*Case 2:* $X = t < 0$, $F = 0$, $U > U_c^{(2)}$, for a lattice with loops. (In the case of a bipartite lattice, a phase transformation on every other site may be applied to include the case $t > 0$.) We know that $|\psi_0\rangle$ is a ground state of $\hat{H}_{NN}$ for $\tilde{U} > 0$ and that it is lower in energy than any state with doubly occupied sites. Hence no ground state of $\hat{H}_{NN}$ can have doubly occupied sites. Among the states without double occupancies those with $S = S_{max}$ are the lowest in energy; this follows from the $P$-term in (A1) by a proof completely analogous to that of Tasaki.[5] In Nagaoka's basis,[4]

$$|i, (\sigma_1 \ldots \sigma_{i-1}\sigma_{i+1} \cdots \sigma_L)\rangle = (-1)^i \hat{c}^+_{1,\sigma_1} \cdots \hat{c}^+_{i-1,\sigma_{i-1}} \hat{c}^+_{i+1,\sigma_{i+1}} \cdots \hat{c}^+_{L,\sigma_L}|0\rangle,$$

all off-diagonal matrix elements of the $P$-term are negative. Furthermore, it is represented by an irreducible matrix since it is assumed that the lattice allows the hole to move around loops. It follows directly from the Perron-Frobenius theorem[30] that, in every sector with fixed $\hat{S}^z$, the ground state is unique and is given by a linear combination with strictly positive coefficients of the basis vectors. These ground states are just the states $\{|\psi_M\rangle\}$ of eq. (6) and all have $S = S_{max}$.